\documentclass[a4paper, twocolumn, 10pt, nopdfoutputerror]{revtex4-2}

\usepackage[T1]{fontenc}
\usepackage{amsmath}
\usepackage{amssymb}
\usepackage{empheq}
\usepackage{enumitem}
\usepackage{graphicx}
\usepackage{svg}
\usepackage{booktabs}
\usepackage[braket, qm]{qcircuit}
\usepackage{braket}
\usepackage{hyperref}
\usepackage{subcaption}
\usepackage{placeins}
\usepackage{xcolor}
\usepackage{soul}
\usepackage{tikz}
\usepackage{bbm}
\hypersetup{
	colorlinks   = true, 
	urlcolor     = blue, 
	linkcolor    = blue, 
	citecolor   = red 
}

\newcommand{\var}{\mathbb{V}}
\newcommand{\ex}{\mathbb{E}}
\newcommand{\BP}{BP}
\newcommand{\CP}{Clifford}

\newcommand{\phic}{\phi_{\mathcal{C}}}

\newtheorem{theorem}{Theorem}
\newtheorem{lemma}{Lemma}

\begin{document}
\title{Barren plateaus are swamped with traps}
\begin{abstract}
Two main challenges preventing efficient training of variational quantum algorithms and quantum machine learning models are local minima and barren plateaus. Typically, barren plateaus are associated with deep circuits, while shallow circuits have been shown to suffer from suboptimal local minima. We point out a simple mechanism that creates exponentially many poor local minima specifically in the barren plateau regime. These local minima are trivial solutions, optimizing only a few terms in the loss function, leaving the rest on their barren plateaus. More precisely, we show the existence of approximate local minima, optimizing a single loss term, and conjecture the existence of exact local minima, optimizing only a logarithmic fraction of all loss function terms. One implication of our findings is that simply yielding large gradients is not sufficient to render an initialization strategy a meaningful solution to the barren plateau problem.
\end{abstract}

\author{Nikita A. Nemkov}\email{nnemkov@gmail.com}
\affiliation{National University of Science and Technology ``MISIS”, Moscow 119049, Russia}
\author{Evgeniy O. Kiktenko}
\affiliation{National University of Science and Technology ``MISIS”, Moscow 119049, Russia}
\author{Aleksey K. Fedorov}\email{akf@rqc.ru}
\affiliation{National University of Science and Technology ``MISIS”, Moscow 119049, Russia}
\maketitle
	
\section{Introduction} \label{sec intro}
Variational quantum algorithms (VQAs)~\cite{Cerezo2021} and a closely related field of quantum machine learning (QML)~\cite{Biamonte2017, Schuld2014} are still the leading paradigms for quantum computing in the NISQ era \cite{Preskill2018, Bharti2021,Fedorov2022}. They are considered a natural generalization of the classical neural networks, with the potential to leverage the power of quantum information processing. While VQAs and QML are expected to surpass classical models in certain aspects, only a few theoretical results concerning their performance are available up to date.

One of the key obstacles facing VQAs and QML models~\cite{Thanasilp2021, Rudolph2023a} is trainability. Loss functions  in classical deep neural networks are known to be highly non-convex~\cite{Goodfellow2016}. Nevertheless, they usually can be optimized with surprising efficiency by gradient-based methods, i.e. they are \textit{trainable}. Unfortunately, the trainability of VQAs appears to be a qualitatively harder problem. Two main issues are poor \textit{local minima} (LM) and \textit{barren plateaus} (BPs). It is possible to construct VQAs that are provably NP-hard to optimize owing to the exponential number of LM in the loss landscape~\cite{Bittel2021}. More importantly, the domination of poor LM appears to be a generic~\cite{Anschuetz2021, Anschuetz2022}, rather than a fine-tuned~\cite{Bittel2021, You2021} property.

BPs, as first pointed out in Ref.~\cite{McClean2018} and elaborated in many subsequent works (see Ref.~\cite{Larocca2024} for a succinct recent review), describe the situation where the loss function has exponentially small (in the number of qubits $n$) variance. In other words, the loss landscape is basically flat with exponential precision everywhere, possibly except for a parameter region of an exponentially small volume. The presence of BPs is a generic property, manifesting the curse of dimensionality associated with the exponentially large Hilbert space. The most important cause of BPs is \textit{expressivity} of a VQA, which generically appears for sufficiently deep $poly(n)$ circuits. Hence, BPs are often considered to be a problem of deep circuits. In contrast, shallow $poly(log(n))$ quantum circuits can be free of BPs, but have been shown to suffer from poor LM~\cite{Anschuetz2021, Anschuetz2022}.

\begin{figure}
	\includegraphics[width=0.5\textwidth]{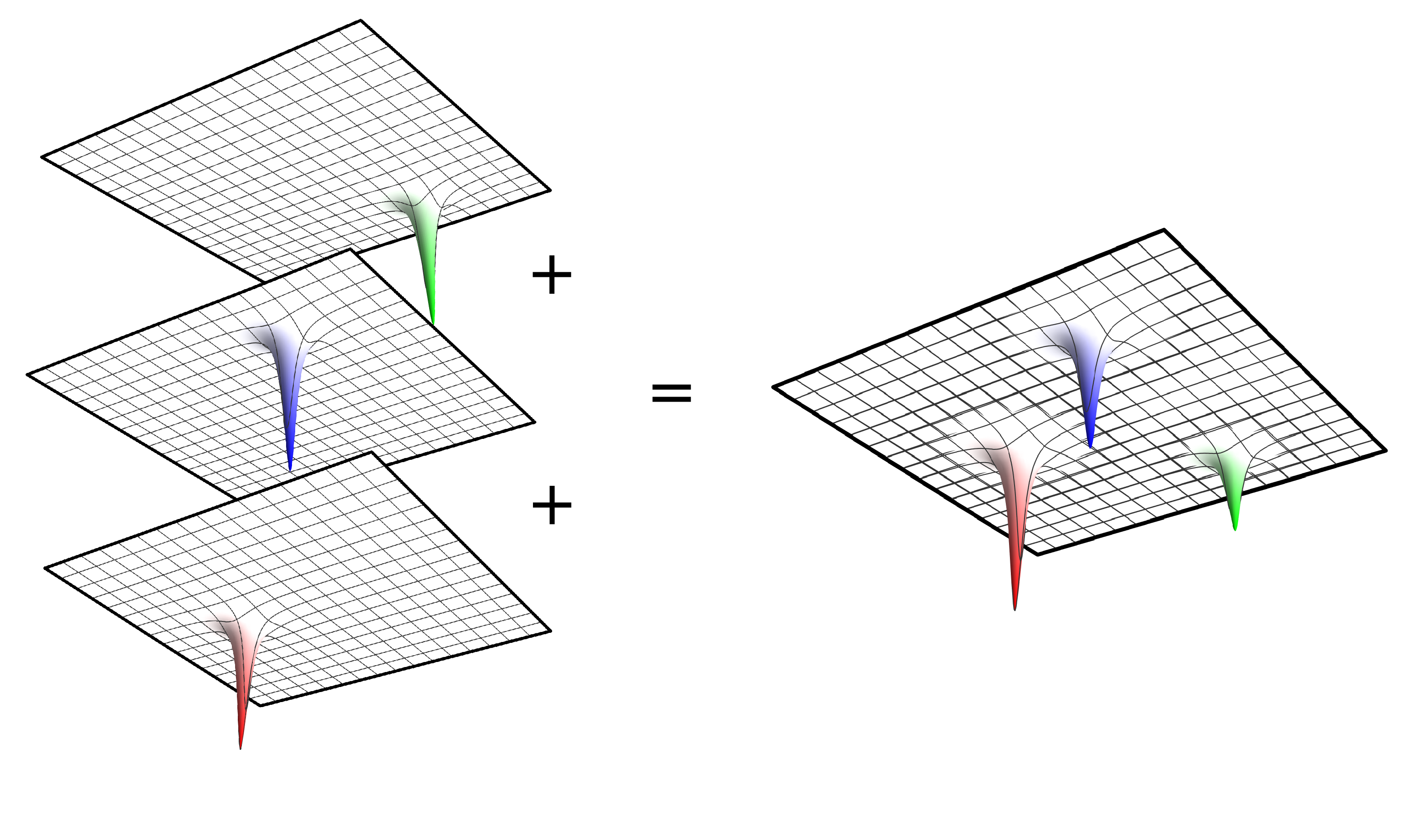}
	\caption{Schematic illustration of the main idea. When the loss functions corresponding to the individual observables are highly concentrated (left), their respective local minima are likely to remain the local minima of the total loss function (right).}
	\label{fig loss}
\end{figure}

Here we argue that a very general class of VQAs contains exponentially many LM of a particular type in the BP regime as well. The very possibility of LM coexisting with BP is not new. Theoretically \cite{Larocca2021, Anschuetz2021, Anschuetz2022} and empirically \cite{Campos2021a, Campos2021b, Kiani2020, Nemkov2023b}, LM have been shown to disappear only in the in overparameterized circuits, which typically implies exponential depth far beyond the onset of BPs. However, the structure of these LM is not well-understood, for instance \cite{Anschuetz2021} suggested that they might be mostly represented by exponentially shallow bumps. In contrast, here we show that there can be exponentially many LM with sufficiently large (absolute) loss values, accompanied by sufficiently large gradients in their vicinity.

The basic idea is simple and can be sketched as follows. An observable $O$ in a VQA is typically a sum of simpler operators $O=\sum_i O_i$.
To each of these we can associate a separate loss function $L_i(\pmb\phi)$, so that the full loss function is their sum $L(\pmb\phi)=\sum_i L_i(\pmb\phi)$, where $\pmb\phi$ is a vector of VQA's parameters. If the full loss function $L(\pmb\phi)$ has a BP, each of the contributing terms $L_i(\pmb\phi)$ has a BP as well. 

Then, any point $\pmb\phi_i$ that is a local minimum for some \textit{particular} $L_i$, is likely to be a local minimum for the \textit{total} loss function $L$ as well, simply because the contribution of the other terms $L_{j\neq i}$ is exponentially close to zero over most of the parameter space, see Fig.~\ref{fig loss} for an illustration. Typically, a single term $L_i$ captures only a tiny bit of information about the original problem, e.g. a single interaction term in a molecular Hamiltonian or a single constraint in a combinatorial optimization problem. Hence, such points, which we refer to as \textit{siloed} LM, are expected to yield very poor solutions.

To make this idea precise, we need to (i) be able to characterize LM of individual loss functions $L_i$ and (ii) prove that LM of different terms are sufficiently uncorrelated to not interfere with each other. In this work, we focus on a class of VQA we refer to as \textit{Clifford VQAs}, which only involve constant Clifford gates, Pauli observables, and parameterized Pauli rotations. Clifford VQAs provide us with sufficient control to make a precise argument and at the same time, they capture most of ansatz structures appearing in practice such as QAOA~\cite{Farhi2014}, the majority of hardware-efficient circuits~\cite{Kandala2017} and VQE~\cite{Romero2018}. We also discuss whether our conclusions are likely to apply to other types of VQAs in Sec.~\ref{sec disc}.

In practice, the behavior depicted in Fig.~\ref{fig loss}, gives an oversimplified picture in most cases. In the Clifford VQA setting, we find that the LM of individual Pauli terms are not isolated points, but rather high-dimensional surfaces, as illustrated in Fig.~\ref{fig loss gorges}, and discussed in detail later. We will show that points on these critical surfaces typically are approximate LM (up to exponentially vanishing gradients), and argue that intersection of ${\cal O}(\log(n))$ such surfaces are exact LM. However, at the beginning, we encourage the reader to refer to the more intuitive picture sketched in Fig.~\ref{fig loss}.

Our findings add to the bundle of challenges that need to be addressed when designing and training VQAs. There are two primary ways to deal with the BP problem. The first is to use ansatz structures that are free of BPs by design, such as shallow local circuits~\cite{Pesah2021}, circuits with polynomial dynamical Lie algebras~\cite{Larocca2021, Larocca2021a} or special symmetry structure~\cite{Larocca2022, Meyer2022}. However, a growing body of evidence, distilled in recent work~\cite{Cerezo2023}, implies that the very same properties that prevent BPs make the relevant VQAs classically simulable. Another approach is to use specific parameter initializations that, while not eliminating the BPs, prevent the training from starting there. However, most of these initialization strategies rely on the initial quantum circuit being close to the identity~\cite{Skolik2020, Grant2019, Zhang2022d, Wang2023}. It is not clear a priori whether such initializations allow the optimization to cross over into the deep circuit regime. Other initialization strategies exits that target deeps circuits~\cite{Rudolph2022}, or even specifically Clifford circuits~\cite{Ravi2022}. However, in all those cases, the possibility to end up in a poor LM should not be ignored. Hence, we argue that an initialization leading to large gradients is not synonymous with solving the BP problem.

\section{Barren plateaus} \label{sec bp}
\subsection{Definition and origins}
A VQA is defined by an initial state $\rho$, a parameterized quantum circuit $U(\pmb\phi)$, and an observable $O$, with the loss function given by the expectation value
\begin{align}
L(\pmb \phi)=\operatorname{Tr}\left[\rho\, U^\dagger(\pmb \phi )\,O\, U(\pmb \phi)\right] \ , \label{loss def}
\end{align}
with $\pmb\phi=(\phi_k)$ being a vector of parameters. The loss function $L(\pmb\phi)$ is said to have a \BP{} if its values are exponentially concentrated around the mean $\ex [L]$, i.e. when the variance of the loss function, defined by
\begin{align}
\var [L]=\ex\left[\left(L(\pmb\phi)-\ex [L]\right)^2\right] \ , \label{var loss}
\end{align}
vanishes exponentially with the number of qubits $\var[L]={\cal O}(b^{-n})$ for some $b>1$.
In the standard setting $b=2$, which is directly related to the dimension of the $n$-qubit Hilbert space. For functions $F(\pmb\phi)$ periodic with respect to each variable $\phi_k$, which is the most common case in practice, expectation values are calculated as $\ex [F]=\int_0^{2\pi} \prod_{k}\frac{d\phi_k}{2\pi} \,\,F(\pmb\phi)$.

We note that though originally BPs were defined by vanishing gradients rather than the concentration of loss function~\cite{McClean2018}, these  conditions are in fact equivalent~\cite{Arrasmith2021, Miao2024}, and loss concentration is usually a more convenient diagnostic tool. Still, some of our arguments will rely on the gradient concentration. For a loss function $L(\pmb\phi)$ with variance $\var[L]=\mathcal{O}(b^{-n})$ and for any $\epsilon$, the concentration result can be stated formally using Chebyshev's inequality
\begin{align}
\Pr\left[\Big|L-\ex[L]\Big|>\epsilon\right]\le \mathcal{O}\left(\epsilon^{-2} b^{-n}\right)    \ . \label{concentration}
\end{align}

Four main sources of BPs have been identified in the literature.
\begin{itemize}
	\item \textit{Circuit expressivity.} When the parameterized quantum circuit defining the VQA furnishes an approximate unitary 2-design, $L$ suffers from a BP~\cite{McClean2018, Holmes2021}. Typically, local circuits beyond linear depth $d={\cal O}(n)$ are sufficient to qualify for 2-designs~\cite{Dankert2006, Harrow2018a}, while the logarithmic depth $d={\cal O}(\log n)$ circuits are not.
	\item \textit{Non-locality of the observable.} VQAs with shallow local circuits can have a BP if the observable $O$ is sufficiently non-local \cite{Cerezo2021a, Uvarov2020a}, which is the case in many of the proposed VQA applications such as, for example, quantum compiling~\cite{Khatri2018, Jones2018b} or autoencoders~\cite{Romero2016}.
	\item \textit{Entanglement of the initial state}. BP can arise in VQAs with sufficiently entangled initial states $\rho$, which is common in quantum machine learning and quantum chemistry~\cite{OrtizMarrero2021, Thanasilp2021, Sharma2020,Sapova2022}.
	\item \textit{Noise}. Sufficiently strong noise also generically leads to BPs~\cite{Wang2020, Franca2020, Fontana2023, Singkanipa2024}.
    Noise-induced BPs are qualitatively different from the rest, as the noise tends to flatten the loss landscape completely, instead of concentrating good solutions in a small parameter region. 
	\end{itemize}
 
Interestingly, our argument will be sufficiently general and apply to any source of the BP (except for the noise), so that the classification above will be useful to interpret the results, but not necessary to derive them. At the same time, we will heavily rely on the properties of a specific class of VQAs, that we now introduce.

\subsection{\CP{} VQAs} \label{sec cvqa}
The class of VQAs we consider here restricts the initial state $\rho$ to be a stabilizer state, the observable $O$ to be a sum of $poly(n)$ Pauli stings $O=\sum_i c_i P_i$ with real coefficients $c_i$, and the circuit $U(\pmb\phi)$ to consist only of the constant Clifford gates and parametric Pauli rotation gates $e^{-\imath P_k \phi_k/2}$. Generators of the parametric gates $P_k$ and Pauli observables $P_i$ may or may not be related. We will refer to such VQAs as \textit{\CP{} VQAs}. We also assume that there are no correlated parameters, i.e. that all angles $\phi_k$ are independent, and that there are at most $poly(n)$ parameters in total.

Denote by $|\pmb \phi|$ the total number of parameters in the circuit, by $\pmb\phic$ a discrete set $\pmb\phic=\{0, \pi/2, \pi, 3\pi/2\}^{|\pmb\phi|}$ (the set of all $4^{|\pmb \phi|}$ combinations obtained by setting each component of $\pmb\phi$ to either $0, \pi/2, \pi$ or $3\pi/2$). 
The key technical fact allowing to derive most of our quantitative results is the following

\begin{lemma}[Averaging over Clifford points]\label{lemma avg}
	Let $F(\pmb\phi)$ be a trigonometric polynomial of degree at most two with respect to each angle $\phi_k$. 
    Then,
	\begin{align}
	\ex [F] = \frac{1}{4^{|\pmb\phi|}}\sum_{\pmb\phi_c\in \pmb\phic} F(\pmb\phi_c) \ .
	\end{align}
\end{lemma}
The simple proof, as well as a precise definition of degree two trigonometric polynomial, is found in App.~\ref{app lemma}. With our definition of a \CP{} VQA, and in the absence of correlated parameters, both $L(\pmb\phi)$ and $L^2(\pmb\phi)$ satisfy conditions of this Lemma, a simple fact made explicit in App.~\ref{app trig}.
This allows expressing variance~\eqref{var loss} as the sum over the discrete set of points $\pmb\phic$. For $\pmb\phi_c\in\pmb\phic$ all the Pauli rotation gates in $U(\pmb\phi_c)$ become Clifford operators and $U(\pmb\phi_c)$ becomes a Clifford circuit, hence we refer to $\pmb\phic$ as the \textit{Clifford points}. 

Now let us consider the case where the observable is a single Pauli string $P$. Then, the value of the loss function at a Clifford point $\pmb\phi_c$ can be either $\pm1$ or $0$. This is because at a Clifford point $U^\dagger(\pmb\phi_c)P U(\pmb\phi_c)$ is again a Pauli string, and its expectation value in the stabilizer state $\rho$ is either $0$ or $\pm1$. This basic fact allows us to derive the following
\begin{theorem}[Concentration at Clifford points] \label{thm zero}
	Let the loss function $L$ of a \CP{} VQA have a BP with $\var[L]={\cal O}(b^{-n})$. Then, for a randomly sampled Clifford point $\pmb\phi_c\in\pmb\phic$, the probability that $L(\pmb\phi_c)$ is not equal to its expected value vanishes exponentially
	\begin{align}
	\Pr(|L(\pmb\phi_c)-\ex[L]|>0)={\cal O}(b^{-n})\ . \label{prob l}
	\end{align}
	Similarly, any gradient component or Hessian entry is zero with probability exponentially close to one
	\begin{align}
	&\Pr(|\nabla_{\phi_k}L(\pmb\phi_c)|>0) = {\cal O}(b^{-n}),	\label{prob gl}\\ &\Pr(|\nabla_{\phi_k}\nabla_{\phi_l}L(\pmb\phi_c)|>0) = {\cal O}(b^{-n}) \ . \label{prob hl}
	\end{align}
\end{theorem}
The proof if straightforward. For simplicity, and without the loss of generality, we set $\ex[L]=0$, so that the variance is given by $\var[L]=\ex[L^2]$ (see App.~\ref{app avg} for justification of this assumption). First, consider the case of a single Pauli observable $P_i$. As explained above, the value of the corresponding loss function at the Clifford points is either $L_i^2=1$ or $L_i^2=0$. Therefore, according to Lemma~\ref{lemma avg}, the variance
\begin{align}
    \var[L_i]=\frac{1}{4^{|\pmb\phi|}}\sum_{\pmb\phi_c\in \pmb\phic} L_i^2(\pmb\phi_c) \label{var as sum}
\end{align}
simply counts the number of non-zero values at Clifford points normalized by the total number of Clifford points. If $L_i$ has a BP, this variance, and hence the proportion of the Clifford points with non-zero value, is exponentially small, leading to~\eqref{prob l}. Clearly, replacing a single Pauli observable $P_i$ by a $poly(n)$ sum of Pauli strings $O=\sum_i c_i P_i$ can not overcome the exponential concentration. The reasoning extends to the derivatives as well. For instance, using the parameter-shift rule~\cite{Mitarai2018, Schuld2018} the gradient can be expressed as
\begin{align}
\nabla_{\phi_k}L(\pmb\phi)=\frac12L\left(\pmb\phi+\frac{\pi}{2}\pmb e_k\right)-\frac12L\left(\pmb\phi-\frac{\pi}{2}\pmb e_k\right) \ , \label{shift}
\end{align}
where $\pmb e_k$ is a unit vector in the direction of $\phi_k$. Each of the two terms contributing to the gradient (i) has a BP (ii) assumes a discrete set of values $\pm\frac12, 0$ at Clifford points and (iii) satisfies conditions of Lemma~\ref{lemma avg}. Hence, their sum will inherit all these properties and then Eq.~\eqref{prob gl} can be deduced in the same way as Eq.~\eqref{prob l} (note that $\ex[\nabla_{\phi_k} L]=0$ is manifest from \eqref{shift}). Similar reasoning applies to the Hessian components, and we will not spell it out.

The main conclusion of this section is as follows. Almost by definition, randomly sampling the parameters of a VQA suffering from a BP leads to exponentially small loss value, with probability exponentially close to one. We have shown that for a \CP{} VQA random sampling of a Clifford point leads to loss being \textit{exactly} zero, with probability exponentially close to one. Moreover, the same applies to any finite derivative of the loss function. 
\section{Traps in barren plateaus} \label{sec traps}

\subsection{Loss landscapes of single Pauli terms} \label{sec loss single}
\begin{figure}
    \centering
    \includegraphics[width=0.5\textwidth] {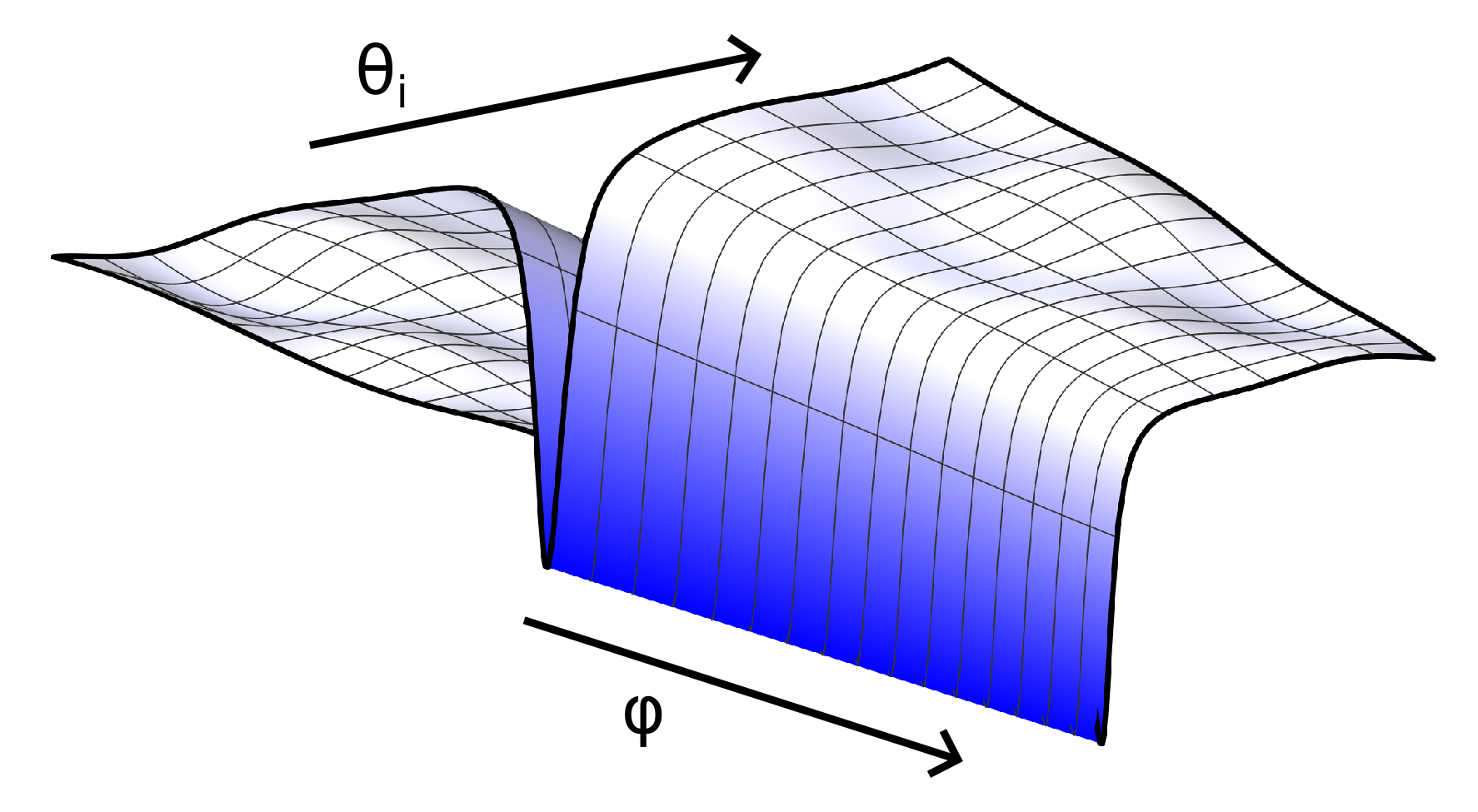} 
    \caption{Local structure of the critical surface for a single Pauli observable.}
    \label{fig gorge}
\end{figure}

We begin by discussing the structure of the loss landscapes $L_i$, involving a single Pauli observable $P_i$. One simple observation is that any point $\pmb\phi_i$ yielding $L_i(\pmb\phi_i)=-|c_i|$ is a global minimum of $L_i(\pmb\phi)$, because the expectation value of $P_i$ is bounded between $-1$ and $+1$. Another point, which is less obvious, is that LM of $L_i$ almost never are isolated points, but rather hypersurfaces of high dimension. In other words, the parameter vector $\pmb\phi_i$ typically admits a split $\pmb\phi_i$ into fixed $\pmb\theta_i$ and free $\pmb\varphi$ components 
\begin{align}
\pmb\phi_i=(\pmb\theta_i, \pmb\varphi),   
\end{align}
such that changing any of the parameters $\pmb\theta_i$ leads away from the critical surface, while changing $\pmb\varphi$ has no effect, i.e. $L_i(\pmb\theta_i,\pmb\varphi)=-|c_i|$ for any choice of $\pmb\varphi$ (not necessarily Clifford). This is illustrated in Fig.~\ref{fig gorge}. We will refer to $\pmb\varphi$ as null directions. Note that $L_i$ needs to be independent of $\pmb\varphi$ only when $\pmb\theta_i$ are fixed, i.e. only the bottom of the gorge needs to be flat. 

An easy case showing that null directions exist is a VQA with local observables. Then, all the parameters in $\pmb\phi_i$, whose generators lie outside the observable's lightcone, are trivially null directions. However, null directions appear much more generally. To see this, fix all parameters $\pmb\phi$ to their values implied by $\pmb\phi_i$, except for a single angle $\phi_k$. As a function of $\phi_k$, the circuit has the form $U(\phi_k) = C_L e^{-i P_k \phi_k/2}C_R$ with both $C_L$ and $C_R$ being Clifford gates (which are combinations of the original Clifford gates, and Clifford gates resulted from fixing all the other parameterized gates to their Clifford values). The loss function then is
\begin{align}
L_i(\phi_k)=\operatorname{Tr}\left[\rho\,\, C_R^\dagger e^{iP_k\phi_k/2} C_L^\dagger P_i C_L e^{-iP_k\phi_k/2}C_R\right] \ . \label{null direction}
\end{align}
If the generator $P_k$ commutes with the Pauli operator $C_L^\dagger P_i C_L$, then $L_i(\phi_k)$ in fact does not depend on $\phi_k$ and hence $\phi_k$ is a null direction. If $P_k$ is outside of $P_i$'s lightcone, these operators necessarily commute. However, even when the lightcones of $P_i$ and $P_k$ intersect, we expect them to commute roughly half of the times, simply because two random Pauli strings sharing qubits have one half probability of commuting. Hence, in a generic case, we expect all the parameters outside the lightcone of $P_i$, and roughly half of the parameters within, to be null directions for $L_i$, so that $|\pmb\varphi|\ge |\pmb\phi|/2$. As shown in App.~\ref{app null} it is possible to construct loss functions $L_i$ that have few to no null directions at their critical points, but such examples are arguably contrived.

\begin{figure}
    \centering
    \includegraphics[width=0.5\textwidth]{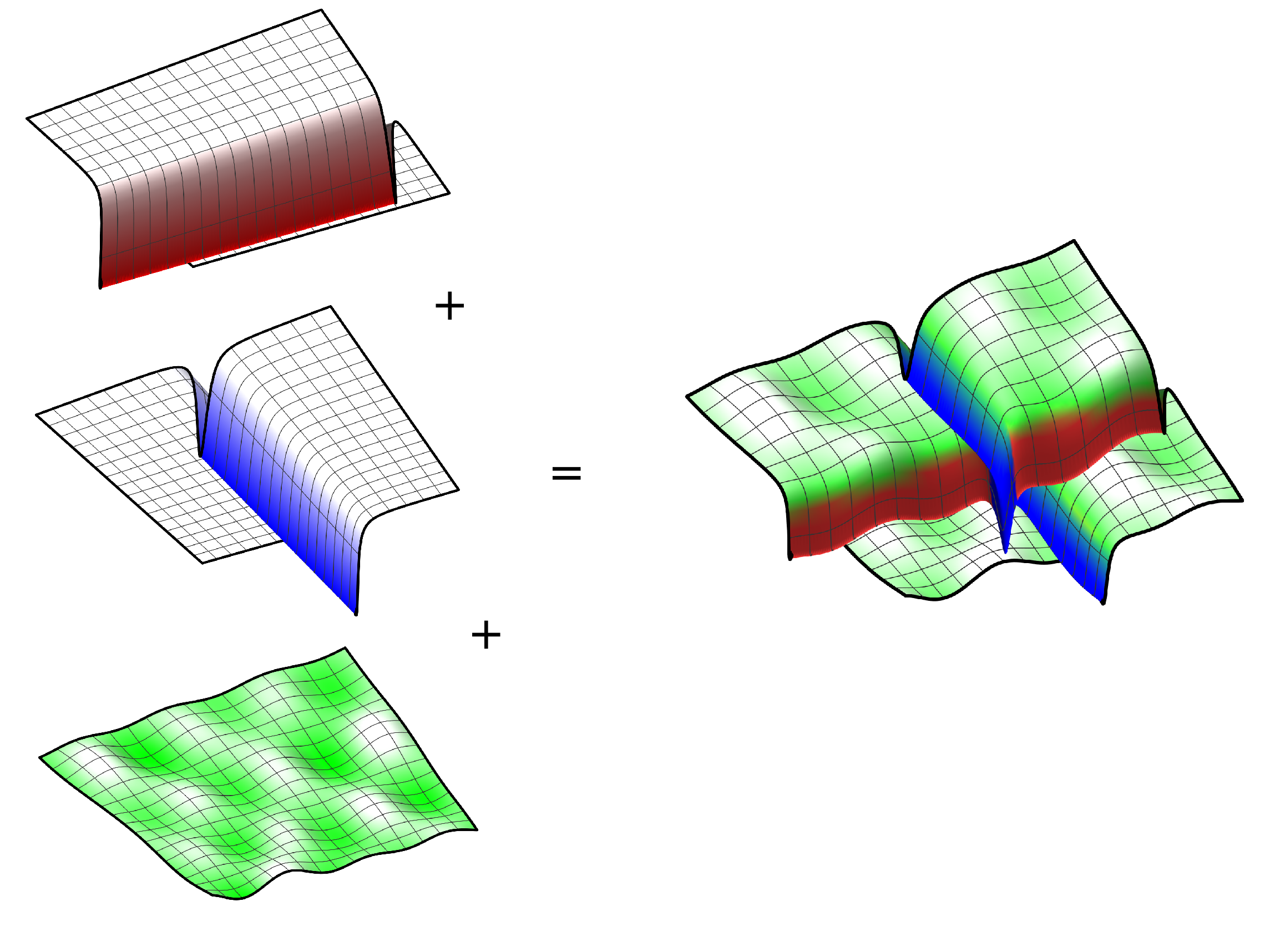}
    \caption{Loss gorges corresponding to individual Pauli terms, and their interaction.}
    \label{fig loss gorges}
\end{figure}

Therefore, we need to update our initial sketch in Fig.~\ref{fig loss} replacing localized sharp peaks with narrow gorges, see Fig.~\ref{fig loss gorges}. This figure suggests the existence of two types of LM, those that belong to a single gorge and optimize a single term in the loss function, and those that lie on the intersection of several gorges and optimize several terms. We will discuss both types in turn.

\subsection{Approximate local minima} \label{sec approximate}

We are now ready to quantitatively argue that quite generally, a random point $\pmb\phi_i$, which is a global minimum for one of the Pauli observables $L_i$, is exponentially likely to be an approximate local minimum for the full loss function $L$. The intuitive definition of an approximate local minimum $\pmb\phi_*$ is that of a point, where it is not possible to meaningfully improve the value of the loss function $L(\pmb\phi_*)$ by any small perturbation. More precisely, following the mathematical literature (see e.g. \cite{Agarwal2016}) we say that $\pmb\phi_*$ is an $\epsilon$-approximate critical point of $L(\pmb\phi)$, if all gradient components at $\pmb\phi_*$ are bounded by $|\nabla_{\phi_k} L(\pmb\phi_*)|\le \epsilon$. If, furthermore, the smallest eigenvalue $\lambda_{\rm min}$ of the Hessian $\nabla^2 L(\pmb\phi_*)$ satisfies $\lambda_{\min}\ge -\sqrt{\epsilon}$, we call $\pmb\phi_*$ an $\epsilon$-approximate local minimum. 

\begin{figure}
    \centering
    \includegraphics[width=0.5\textwidth] {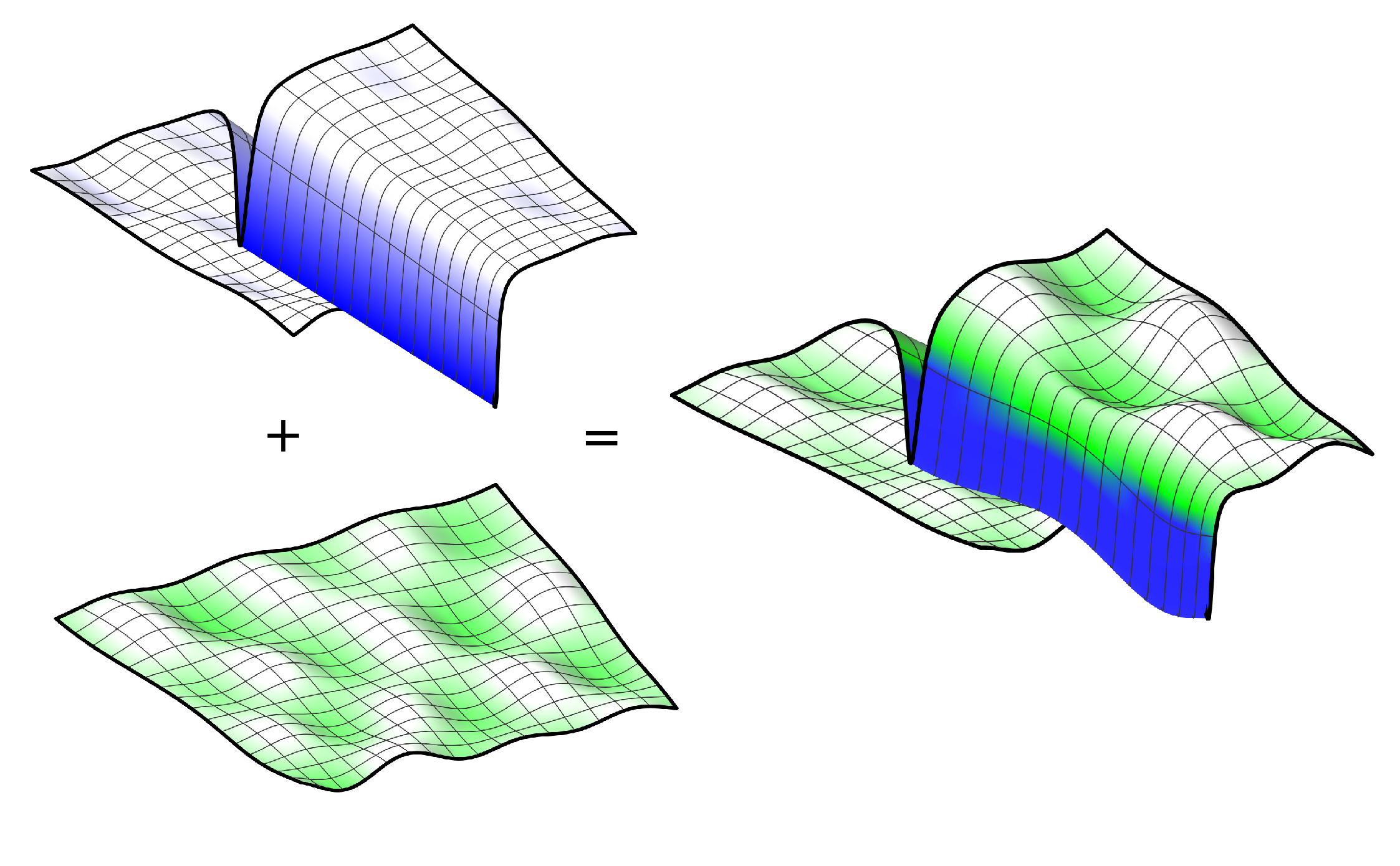} 
    \caption{Local structure of the critical surface for the full loss function.}
    \label{fig gorge bp}
\end{figure}

Our argument is based on a simple variation of the idea presented in the introduction. Let us divide the loss function into $L_i$, the term that was specifically optimized by the choice of $\pmb\phi_i$, and the rest of the terms $R_i(\pmb\phi)=L(\pmb\phi)-L_i(\pmb\phi)=\sum_{j\neq i}L_j(\pmb\phi)$. Assuming that $R_i(\pmb\phi)$ is in some sense independent of $L_i(\pmb\phi)$, the vicinity of the point $\pmb\phi_i$ is likely to be a generic region for $R_i$, and lie on its BP. Hence, adding $R_i$ could only change the value of $L_i$ by an exponentially small margin, and at most turn $\pmb\phi_i$ into an $\epsilon$-approximate local minimum instead of an exact one. This is illustrated in Fig.~\ref{fig gorge bp}. The technical condition leading to the desired notion of independence between $L_i$ and $R_i$ is the requirement that $R_i(\pmb\theta_i,\pmb\varphi)$ has a BP with respect to the null directions of $L_i$. This is formalized by

\begin{theorem}[Approximate local minima] \label{thm minima}
	Let $L(\pmb\phi)=\sum_j L_j(\pmb\phi)$ be a loss function of a \CP{} VQA having a BP, and a Clifford point $\pmb\phi_i$ be a local minimum of some $L_i$ with $L_i(\pmb\phi_i)=-|c_i|$. Denote the rest of the terms in the loss function by $R_i(\pmb\phi)=L(\pmb\phi)-L_i(\pmb\phi)$. Assume that
	\begin{enumerate}
		\item Parameters $\pmb\phi_i$ admit a split $\pmb\phi_i=(\pmb\theta_i, \pmb\varphi)$ such that $L_i(\pmb\theta_i,\pmb\varphi)=-|c_i|$ for any $\pmb\varphi$.
		\item The rest of the terms in the loss function $R_i(\pmb\theta_i,\pmb\varphi)$ have a BP with respect to angles $\pmb\varphi$.
        \item Changing any single fixed angle, or a pair of fixed angles $\pmb\theta_i\to \pmb\theta_i'$, does not remove a BP of $R_i(\pmb\theta_i', \pmb\varphi)$.
	\end{enumerate}
 
(i) Then, a point $\pmb\varphi$ sampled uniformly at random yields an $\epsilon$-approximate local minimum $(\pmb\theta_i,\pmb\varphi)$ of the full loss function with probability $1-\mathcal{O}(\epsilon^{-2}b^{-n})$.

 (ii) Moreover, a point $\pmb\varphi_c$ sampled uniformly from the set of Clifford points $\pmb\varphi_{\mathcal{C}}$, yields a 0-approximate local minimum $(\pmb\theta_i, \pmb \varphi_{c})$ of the full loss function with probability $1-\mathcal{O}(b^{-n})$.
\end{theorem}

The first statement of the theorem is almost trivial. Note that for any loss function $L(\pmb\phi)$ with a BP, a randomly sampled point $\pmb\phi$ is an $\epsilon$-approximate local minimum with probability $1-\mathcal{O}(\epsilon^{-2}b^{-n})$ (choosing $\epsilon$ e.g. as $\epsilon=b^{-n/4}$ shows that a uniformly sampled point is an exponentially precise LM with probability exponentially close to one). This follows from the exponential concentration of the gradient (and Hessian) components, which satisfy concentration inequalities identical to \eqref{concentration}. Hence, the assumptions of the theorem effectively shift the question from whether we expect $R_i(\pmb\phi)$ to have exponentially small values around the  of $L_i$ to whether a constrained version of $R_i(\pmb\phi)$, $R_i(\pmb\theta_i,\pmb\varphi)$ is expected to still have a BP. We will focus on this crucial question in Sec.\ref{plausibility}, but first, let us further clarify and discuss implications of Thm.~\ref{thm minima}.

The assumption of $R_i(\pmb\theta_i,\pmb\varphi)$ having a BP with respect to $\pmb\varphi$ guarantees concentration of the gradients with respect to $\pmb\varphi$. However, it does not necessarily imply that the gradients of $R_i$ with respect to $\pmb\theta_i$ are small. This is why we need the third, somewhat technical assumption, in the theorem. The shift rule \eqref{shift} relates gradients with respect to $\pmb\theta_i$ to the values of $R_i$ with one of $\pmb\theta_i$ shifted, and the assumption that the BP persists under such shifts then guarantees the concentration of the gradients with respect to $\pmb\theta_i$ as well. Similarly, Hessian entries can be related to the values of $R_i$ with two of parameters $\pmb\theta_i$ shifted, and are also concentrated under the assumptions of the theorem. This, in effect, proves the first statement of the theorem.

The second statement of Thm.~\ref{thm minima} follows directly from Thm.~\ref{thm zero}. Namely, if instead of sampling $\pmb\varphi$ uniformly we sample $\pmb\varphi_c$ from the set of Clifford points $\pmb\varphi_{\mathcal{C}}$, any individual gradient or Hessian component is exactly zero with probability exponentially close to one. Since there are at most $poly(n)$ such components, the full gradient vector and Hessian matrix vanish exactly at $(\pmb\theta_i,\pmb\varphi_c)$ with probability exponentially close to one, making it a 0-approximate local minimum. 

We should note that a 0-approximate local minimum $\pmb\phi_*$ need not be a true local minimum of the original function, but only of its quadratic approximation. It is possible that higher-order terms can decrease the value of the function in the vicinity of $\pmb\phi_*$  (a simple example is $f(x)=x^3$ for which $x_*=0$ is a 0-approximate local minimum but not a true local minimum). That said, both exponentially precise and $0$-approximate LM pose a significant practical challenges for optimization. The former can only be escaped along directions with exponentially small gradients, which is exactly the same problem that renders optimization of the original loss function challenging, while the latter are literally indistinguishable from true LM  for any first- or second-order gradient based optimization. We will also discuss \textit{exact} LM in Sec.~\ref{sec exact}.

\subsection{Plausibility and exceptions} \label{plausibility}

Now let us discuss why the assumptions stated in Thm. \ref{thm minima} are expected to hold rather generally, as well as possible exceptions. While the assumptions were intentionally formulated to be ansatz-agnostic, we will need to get problem-specific when discussing their applicability.

In Sec.~\ref{sec loss single} we argued that the LM of individual Pauli terms $\pmb\phi_i$ typically admit a nontrivial split $\pmb\phi_i = (\pmb\theta_i, \pmb\varphi)$ with many null directions $|\pmb\varphi|\sim |\pmb\phi|/2$. Here we point out that it is possible to construct loss functions $L_i$ having few or no null directions at their critical points, but such examples are arguably contrived, see App.~\ref{app null}.

Let us now explain why it is natural for the BP to persist in $R_i(\pmb\phi)$ even after fixing part of the parameters $\pmb\theta_i$. A coarse version of the argument is as follows. First, suppose that fixing parameters $\pmb\theta_i$ effectively removes the source of the BP, whatever its origin. For instance, fixing too many $\pmb\theta_i$ so that $|\pmb\varphi|=O(\log n)$ can reduce the circuit expressivity. Or, in case of an entanglement-induced BP, fixing $\pmb\theta_i$ can lead to a circuit that is undoing the entanglement of the initial state. For a nonlocality-induced BP, one might imagine that all observables $P_i$ can be made local by the same Clifford transformation (i.e. all $C P_i C^\dagger$ are local for some $C$), and fixing $\pmb\theta_i$ effectively performs such a transformation. Though some of these scenarios may appear fine-tuned, they are possible. However, we argue that such choices of $\pmb\theta_i$ are in a sense trivial, and can not lead to interesting solutions. Indeed, by eliminating the very source of BPs these circuit restrictions break the connection to the original problem. For instance, if a shallow VQA is sufficient, there is no point starting with a deep circuit and then eliminating most of the parameters $\pmb\theta_i$ to make the restricted circuit shallow. Or, if the circuit undoes the entanglement of the initial state, it also erases the information encoded there.

We should, therefore, focus on the cases when fixing $\pmb\theta_i$ does not eliminate the source of the BP. But then, the fact that $R_i(\pmb\theta_i,\pmb\varphi)$ still has a BP is almost a direct implication. The only subtlety to address here, is why the BP is present in $R_i$ but absent in $L_i$, the term that is optimized by the choice of $\pmb\phi_i$. (Strictly speaking, $L_i(\pmb\theta_i,\pmb\varphi)$ is constant with respect to $\pmb\varphi$ and hence has a BP, but simply relaxing any of the angles $\pmb\theta_i$ gives a non-constant and non-concentrated loss function).

This tension indeed exists in the common model of BPs. For concreteness, take the expressivity-induced BP. By assumption, the circuit $U(\pmb\theta_i, \pmb\varphi)$ is still highly expressive as a function of $\pmb\varphi$, so we expect BPs in the landscape of any observable. However, to prove that \textit{any} choice of the observable leads to the BP in this scenario, the standard model of expressivity-induced BPs assumes that the underlying circuit furnishes an (approximate) 2-design \cite{McClean2018, Ragone2023, Fontana2023a}.  The 2-design assumption, which in a sense amounts to the uniformity of covering the Hilbert space, is partly a technical convenience. A circuit that explores a large enough portion of the Hilbert space, i.e. is sufficiently expressive, is still expected to produce a BP whether it is a 2-design or not. Without the uniformity guaranteed by the 2-design property, however, \textit{some} observables can give rise to loss functions free of BPs, even with highly expressive circuits. 

We conjecture that this is what happens rather generically. Even if the original circuit $U(\pmb\phi)$ is a 2-design, fixing $\pmb\theta_i$ makes the distribution of $U(\pmb\theta_i, \pmb\varphi)$ non-uniform, biasing it towards one particular observable $P_i$. We anticipate that different models of BPs can be constructed, where the assumptions of being a 2-design are relaxed, at the cost of allowing an exponentially small fraction of the observables to have BP-free loss functions. We discuss one such model for nonlocality-induced BPs in Sec.~\ref{sec random}, and present numerical evidence for expressivity-induced BPs in Sec.~\ref{sec numerics}. In such models, the second assumption of Thm.~\ref{thm minima} will hold with exponentially high probability.

\subsection{A different model of barren plateaus: random Pauli observables} \label{sec random}

Here we introduce a simple scenario meeting all conditions of Thm.~\ref{thm minima}. Consider a Clifford VQA with a single Pauli observable $P$, which is a random Pauli string supported on all qubits (possibly being the identity on some of them). The initial state is $\rho=(|0\rangle\langle 0|)^{\otimes n}$. Denote the corresponding loss function by $L_P(\pmb\phi)$. We show in App.~\ref{app random}, that the variance of the loss function $\var_{\pmb\phi}[L_P(\pmb\phi)]$, \textit{averaged} over the choice of the observable $P$, satisfies
\begin{align}
\ex_{P}[\var_{\pmb\phi}[L_P(\pmb\phi)]] \le \ex_{P}[\ex_{\pmb\phi}[L_P^2(\pmb\phi)]] =  2^{-n} \ .   \label{expvar}
\end{align}
This relation is exact, and holds for any initial state and any quantum circuit, including circuits with few to no parameters (in the latter case, all variances vanish). We view it as a simple model for BPs caused by non-locality, which does not require assumptions on the circuit structure (e.g. the 2-design property). Equation \eqref{expvar} implies that for any given circuit, a random observable has a BP with probability exponentially close to one. However, it does not rule out the existence of observables without BPs. For example, if the circuit is local and shallow, the local observables will be free of BPs. Yet, they constitute an exponentially small fraction among all random observables, consistently with \eqref{expvar}.

Let us now discuss why all assumptions of Thm.~\ref{thm minima} hold in this model. First, let us note that any Clifford minimum for such a VQA has exactly $|\pmb\phi|/2$ null directions on average. To see this, we can revisit the argument around Eq.~\eqref{null direction} and recall that the probability of any single angle $\phi_k$ being a null direction is determined by whether the corresponding generator $P_k$ commutes with $C_L^\dagger P C_L$. Because $P$ is random, this probability is exactly one half for any choice of $P_k$ and $C_L$. 

The second and key assumption of Thm.~\ref{thm minima}, which appears to be difficult to establish in general, also holds naturally in this model. First note that choosing the angles $\pmb\theta_i$ to optimize $L_i$, one effectively constructs a circuit $U(\pmb\theta_i, \pmb\varphi)$, for which $P_i$ belongs to the exponentially small fraction of observables without a BP. However, this fine-tuning for a single observable $P_i$ has no effect on the rest, because relation \eqref{expvar} works for any underlying circuit. Hence, for any $P_{j\neq i}$ drawn at random, the corresponding loss function $L_j(\pmb\theta_i,\pmb\varphi)$ is guaranteed to have a BP with probability exponentially close to one. Exactly the same reasoning applies to the quantum circuit with one or two (or in fact any number) of $\pmb\theta_i$ altered, justifying the final assumption of Thm.~\ref{thm minima}.

We expect that similar models, allowing to relax some structural requirements on VQAs (e.g. featuring 2-designs) at the cost of making the implications probabilistic, can be developed for expressivity- and entanglement-induced BPs as well. We leave this interesting question for future work, but provide numerical experiments in Sec.~\ref{sec numerics} supporting validity of assumptions in Thm.~\ref{thm minima} for expressivity-induced BPs.

\subsection{Exact local minima} \label{sec exact}
There is a natural follow-up question to our discussion of approximate LM from Sec.~\ref{sec approximate}. There, we argued that finding a local minimum of any single term in the loss function $L_i$ is likely to slice the parameter space in roughly equal parts, consisting of the fixed angles and null directions. Along the null directions, the remainder terms in the loss function $R_i(\pmb\theta_i, \pmb\varphi)$ are expected to have BPs. Hence, we should be able to pick another term $L_j$ and find its local minimum along the null directions $\pmb\varphi$, reducing the number of null directions further. After $k$ steps of this procedure, we have a number of loss function terms $L_{i_1}, L_{i_2},\dots$ jointly denoted by $L_I$, a parameter configuration $\pmb\phi_I$ minimizing all of them simultaneously, and the corresponding split $\pmb\phi_I=(\pmb\theta_I,\pmb\varphi)$. The number of null directions is estimated to be $|\pmb\varphi|\sim|\pmb\phi|/2^{k}$. What happens as $k$ approaches $\log_2|\pmb\phi|$, i.e. how does this procedure end?

One possibility is that as the number of free circuit parameters is reduced, one hits a regime where BPs do not exist, and the process can not continue as described. We argue, however, that a more likely scenario is that BPs persist until the very end, but change their character at some point. When the number of parameters is too small, e.g. $|\pmb\varphi|=\mathcal{O}(\log n)$, loss function variance can not be exponentially small, as can be seen e.g. from \eqref{var as sum}. However, it can be \textit{exactly} zero. This is what happens in our model from Sec.~\ref{sec random}, where BPs are present independently of the number of parameters. And this is what we expect to happen generally. After some number of iterations $k\sim\log_2|\pmb\phi|$, the remaining terms $R_I(\pmb\theta_I, \pmb\varphi)=L(\pmb\theta_I, \pmb\varphi)-\sum_{i\in I} L_i(\pmb\theta_I, \pmb\varphi)$ become \textit{exactly} zero as a function of $\pmb\varphi$, as opposed to merely being exponentially concentrated around zero.

If, in addition, the gradient and Hessian of $R_I$ with respect to the fixed angles vanish identically as a function of $\pmb\phi$, i.e. $\nabla_{\pmb\theta_I}R_I(\pmb\theta_I,\pmb\varphi)\equiv0, \nabla_{\pmb\theta_I}^2R_I(\pmb\theta_I,\pmb\varphi)\equiv0$, the point $\pmb\phi_I$ becomes a genuine local minimum for the full loss function. Indeed, in the vicinity of $\pmb\theta_I$ we have the following expansion
\begin{multline}
L(\pmb\theta_I+\delta\pmb\theta,\pmb\varphi)=L_I(\pmb\theta_I+\delta\pmb\theta,\pmb\varphi)+R_I(\pmb\theta_I+\delta\pmb\theta_I,\pmb\varphi)=\\L_I(\pmb\theta_I+\delta\pmb\theta,\pmb\varphi)+\mathcal{O}(\delta\pmb\theta^3) \ .
\end{multline}
Because $\pmb\phi_I$ is a global minimum for $L_I$, it follows that $\pmb\phi_I$ is a local minimum for the total loss function. The assumption of vanishing gradient and Hessian is the direct counterpart of the third assumption of Thm.~\ref{thm minima}, and is expected to hold under the same qualifications. We present numerical evidence supporting this scenario in Sec.~\ref{sec numerics}

In summary, we expect that along with approximate LM optimizing a single Pauli string, there exist exact LM optimizing $\mathcal{O}(\log n)$ Pauli strings simultaneously. The practical significance of such points is not clear. On the one hand, the number of Pauli terms they optimize is still too small to make them interesting solutions. On the other, they are much less generic than approximate LM, and should not contribute much to the existing optimization difficulties. 

\section{Numerical experiments} \label{sec numerics}

\begin{figure}
\setkeys{Gin}{width=\linewidth}
        \begin{subfigure}[t]{0.40\columnwidth}
        \includegraphics{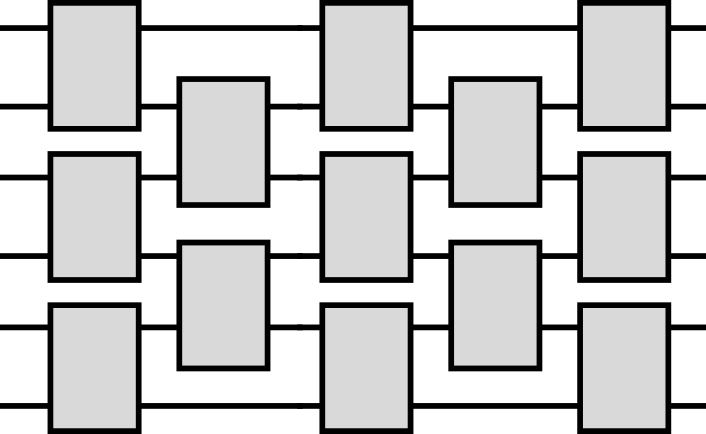}
        \caption{Brickwork circuit architecture. An example with 6 qubits and 5 layers.}
        \label{fig brickwork}
        \end{subfigure}\hfill%
        \begin{subfigure}{0.49\columnwidth}
        \centering
        \includegraphics{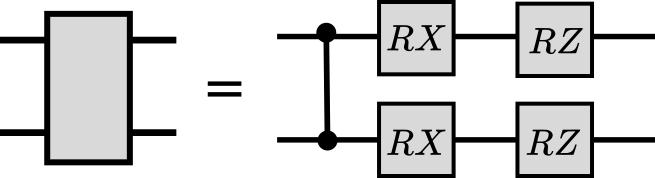}
        \vfill
        \caption{An individual block}
        \label{fig brick}
        \end{subfigure}

        \caption{Circuit structure used in numerical experiments.}
        \label{fig circuit}
\end{figure}

\begin{figure*}
\setkeys{Gin}{width=\linewidth}

        \begin{subfigure}[t]{0.9\columnwidth}
        \includegraphics{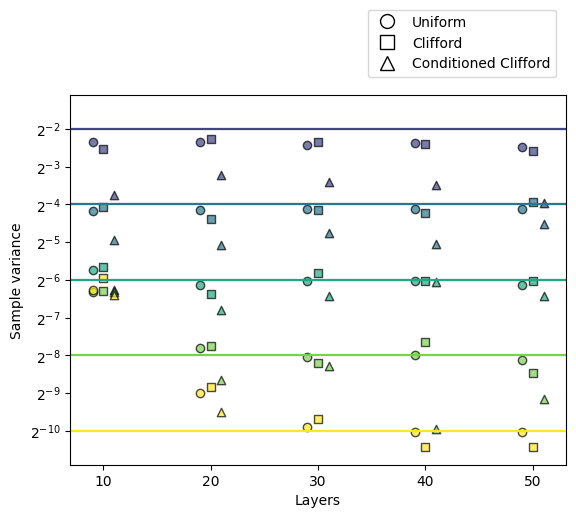}
        \caption{Sample variances of single-Pauli loss functions, averaged over all weight-two nearest-neighbor Pauli strings. Points denoted by $\bigcirc$ are obtained from the uniform sampling of angles, denoted by $\square$ from Clifford sampling. Points denoted by $\triangle$ correspond to statistic over Clifford points, conditioned on having at least one Pauli loss with non-zero value. The number of qubits ($n=2, 4, 6, 8$ or $10$) at each data point is indicated by the color, and coincides with the exponent of the respective horizontal line.}
        \label{fig bp}
        \end{subfigure}\hfill
        \begin{subfigure}[t]{0.9\columnwidth}
        \includegraphics{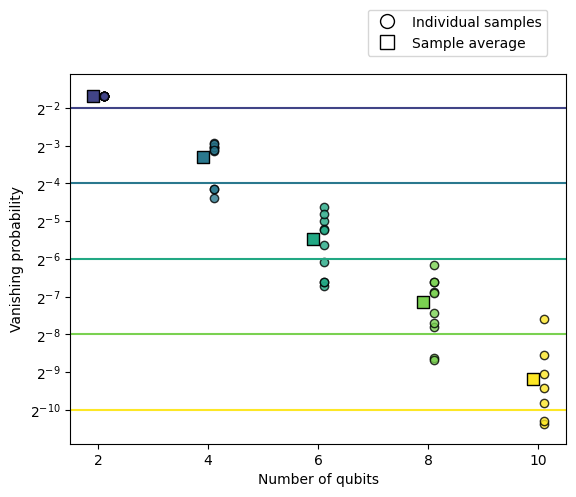}
        \caption{Vanishing probabilities of values and gradients of single-Pauli loss functions, averaged over all weight-two Pauli strings (independent of those defining the candidate exact local minimum). $\bigcirc$ show average vanishing probabilities for individual local minimums (10 points per each value of $n$), $\square$ show averages of all individual points at a given number of qubits.}
        \label{fig exact}
        \end{subfigure}
        
        \caption{Numerical results testing the existence of (a) approximate and (b) exact siloed local minima.}
        \label{fig numerics}

\label{fig experiment}
\end{figure*}

Here, we describe numerical experiments probing our theoretical predictions for the case of expressivity-induced BPs. Generally, numerically quantifying signatures of BPs is challenging, because resolving exponentially small variances requires exponential number of samples. Fortunately, the exponential decay with exponent $2^{-n}$ is expected to hold for small values of $n$ as well (see e.g. \eqref{expvar}, which is valid for any $n$), and allows making quantitative comparisons even with a few-qubit circuits.

In our setup, the circuit has a standard brickwork architecture as depicted in Fig.~\ref{fig brickwork}, where each brick acts on two neighboring qubits and consists of a controlled $Z$ gate followed by an $R_X$ and $R_Z$ rotation on each qubit Fig.~\ref{fig brick}. When enough layers are present, the circuit furnishes a 2-design. Our experiments utilize circuits with different number of qubits and layers, but the general structure is shared by all instances. The initial state is always an all-zero computation basis state $\rho=(|0\rangle\langle 0|)^{\otimes n}$.

Further, we employ two different sets of observables. The first one, denoted by $\mathcal{P}_{2}^{NN}$, consists of all nearest-neighbor Pauli operators of weight two (e.g. $Z_1X_2, Y_3X_4$) and contains $|\mathcal{P}_{2}^{NN}|=9(n-1)$ elements. The second one $\mathcal{P}_{2}$ includes all Pauli strings of weight two, not necessarily nearest-neighbor (e.g. $Z_1X_5$), and has $|\mathcal{P}_{2}|=9n(n-1)/2$ elements.

As a warm-up, we test that uniformly sampling parameters $\pmb\phi$ yields the expected exponential decay of loss variance for deep enough circuits. To this end, we compute sample variance for each Pauli string from $\mathcal{P}_2^{NN}$ and report \textit{average} variance in Fig.~\ref{fig bp}. Results are in good agreement with the expected variance scaling.  We then repeat the same experiment, this time sampling parameters from the set of Clifford points $\pmb\phic$, as in \eqref{var as sum}, and verify that the resulting variances agree with $2^{-n}$ scaling as well. Details of the computational setup, such as the number of samples per observable, are given in App.~\ref{app numerics}.

The next test we perform is non-trivial, and directly checks the validity of the second (independence) assumption of Thm.~\eqref{thm minima} in the current setup, which implies the existence of approximate LM as discussed in Sec.~\ref{sec approximate}. To probe if there are correlations between non-zero values of different Pauli observables, we take the data from the previous step, and only keep the Clifford points $\pmb\phi_c$ which provide non-zero value to at least one of the Pauli observables from $\mathcal{P}_2^{NN}$. We collect the variance statistics over such Clifford points (filtered by having at least one Pauli string with non-vanishing expectation value), and report it in Fig.~\eqref{fig bp}.  There are no signs of positive correlation between non-zero values of the observables, i.e. choosing a Clifford point $\pmb\phi_c$, conditioned only on optimizing a single Pauli observable from $\mathcal{P}_2^{NN}$ but otherwise random, does not seem to bias other Pauli strings towards having non-zero values. 

The second experiment provides a partial test for the existence of exact LM, as described in Sec.~\ref{sec exact}, and proceeds as follows. We sample Clifford points from $\pmb\phic$ at random, and evaluate all weight two Pauli observables from $\mathcal{P}_2$, until we find a Pauli string $P_{i_1}$ with expectation value $-1$. We identify fixed angles $\pmb\theta_{i_1}$ and null directions $\pmb\varphi$ corresponding to this Clifford point. Then, keeping $\pmb\theta_{i_1}$ fixed, we sample Clifford values of null directions $\pmb\varphi$ until finding the next Pauli string $P_{i_2}$ with expectation value $-1$, and update the fixed $\pmb\theta_{i_1i_2}$ and null directions $\pmb\varphi$. We continue this process until no more Pauli strings can be added, obtaining a number of optimized Pauli stings $P_{I}=P_{i_1},P_{i_2},\dots$ and the candidate exact LM $(\pmb\theta_I, \pmb\varphi)$. In practice, we need to impose a threshold on the number of Clifford points to sample, before we consider that adding another Pauli observable is not possible. Details are specified in App.~\ref{app numerics}.

Having a candidate exact LM, our goal is to check if the remaining terms, corresponding to Pauli strings from $\mathcal{P}_2$ not appearing in $P_I$, are exactly zero as a function of $\pmb\varphi$, and so are their gradient (and Hessian) components. To confirm that a loss function $L(\pmb\phi)$ is exactly zero, we sample its values at several uniform points, and verify they are all zero up to a machine precision. For a moderate number of qubits that we work with, the expected variance (about $2^{-10}\approx 10^{-3}$ at worst for $n=10$) is orders of magnitude above machine precision, so we expect this test to be fairly robust. 

A more difficult part is to verify that all gradient components are vanishing. While the number of gradient components scales only polynomially, for a few qubits the variance is not sufficiently small to suppress all of them. For instance, Ref.~\cite{McClean2018} estimated that on a linear topology, depth around $d=10n$ onsets a BP. The corresponding number of parameters (and hence the gradient components) is around $|\pmb\phi|=10n^2$, and one needs to go significantly beyond $n=10$ qubits to make the variance scaling $2^{-n}$ dominant. Instead, we resort to a simpler test, checking the vanishing probability of each gradient component individually.

The final subtlety to our experiment arises from certain peculiarities of the Clifford VQAs at Clifford points. Namely, if two Pauli strings $P_1, P_2$ have non-zero values at some Clifford point $\pmb\phi_c$, so will their product. Indeed, $U(\pmb\phi_c)^\dagger P_1 U(\pmb\phi_c)$ can only have a non-zero expectation value in $\rho=(|0\rangle\langle 0|)^{\otimes n}$ if it consists entirely of $I$ and $Z$ single-qubit Pauli operators, and the same holds for $P_2$. But then their product $U^\dagger(\pmb\phi_c) P_1 P_2 U(\pmb\phi_c)=U^\dagger(\pmb\phi_c) P_1 U(\pmb\phi_c) U^\dagger (\pmb\phi_c)P_2 U(\pmb\phi_c)$ again only contains $I$ and $Z$ factors, and has a non-vanishing value. Therefore, if a Clifford point optimizes Pauli operators $P_1,P_2,\dots$, it also renders the expectation value of any Pauli string generated from them non-zero. Because our Pauli stings come from a set of polynomial (and, in fact, quadratic) size $\mathcal{P}_2$, it is necessary to exclude any Pauli string that can be generated by $P_I$ from $\mathcal{P}_2$ to observe exponential decay of the variance.

With these limitations addressed, we report our numerical results in Fig.~\ref{fig exact}. 
Overall, while the mean vanishing probabilities do not precisely match the anticipated values, the overall decay with exponent around $2$ is clearly visible. We also note that the qualitative behavior of the process described in Sec.~\ref{sec exact} agrees well with numerics. In particular, every additional Pauli generator $P_{i_k}$ added while searching for exact minimum reduces the number of null direction approximately in half, and the total number of optimized Pauli generators $P_I$ is close to the theoretical value $\log_2|\pmb\phi|$ for most trials.

\section{Discussion} \label{sec disc}
In this work, we have argued that loss functions of \CP{} VQAs subject to BPs are likely to contain exponentially many approximate siloed LM, which optimize for a single Pauli observable ignoring all others, as well as exact siloed LM, optimizing only $\mathcal{O}(\log n)$ independent Pauli observables. Apparently, there are many avenues to strengthen and generalize our results, but also many limitations to point out. 

One may wonder whether good solutions are present in the landscape at all. Indeed, given that the LM of individual terms $L_i$ in the loss function are exponentially narrow, and that there is no a priori guarantee that LM of different terms overlap, the siloed LM may be the only solutions in some cases. In other cases, non-siloed LM may exist, but only in the shallow circuit regime. Apparently, making this kind of analysis quantitative requires to examine specifics of the variational ansatz, and we leave this interesting question for future work.

Let us note that our results are compatible with the overparameterization phenomenon in VQA~\cite{Larocca2021, Anschuetz2021, Anschuetz2022}, which suggests that loss landscapes with sufficiently many parameters have most LM exponentially close in value to the global minimum. The necessary number of parameters to achieve overparameterization is of the order of VQA's dynamical Lie algebra dimension, which is exponentially large in a typical BP scenario. So, although by Thm. \ref{thm zero}, the probability of any individual gradient or hessian component being non-zero is still exponentially small, the probability that some non-zero components exists can be of order one in the overparameterized regime.

The key technical limitation of our work is reliance on a specific type of VQAs. However, while the properties of \CP{} VQAs were instrumental to obtain exact results with simple proofs, we expect that the qualitative conclusions may apply more broadly. Apparently, the most essential ingredient for our line of reasoning was the fact that individual Pauli observables have LM with large enough absolute loss values (of order one, as opposed to exponentially small). If a similar property is found in another VQA, we expect it to be accompanied by the siloed LM. 

However, analyzing the loss landscapes of individual terms in general VQAs seems to be a challenging task. Examples include VQA with correlated parameters, non-Clifford gates, or based on non-standard dynamical Lie algebras~\cite{Ragone2023, Fontana2023a}. In all these cases, clarifying the structure of the loss landscape of individual observables may require a challenging and problem-specific analysis. It may happen, in principle, that such deformations of the \CP{} VQA structure spare them from the siloed LM. However, it also seems possible that the resulting landscapes will generally lack any good solutions at all, similarly to the noise-induced BP landscapes. We leave this interesting question for future work.

\section*{Acknowledgments}

N.A.N. thanks the support of the Russian Science Foundation Grant No. 23-71-01095 (obtained theoretical results). Numerical experiments are supported by the Priority 2030 program at the National University of Science and Technology ``MISIS'' under the project K1-2022-027.

\appendix

\section{Trigonometric polynomials}
\subsection{Proof of Lemma~\ref{lemma avg}} \label{app lemma}
We call a single-variable function $F(\phi)$ a trigonometric polynomial of degree $2$ if it admits the following representation
\begin{align}
    F(\phi)=a+b \cos\phi+c\sin\phi+d \cos 2\phi+e\sin 2\phi \ . \label{def deg2}
\end{align}
Note that using identities $\cos 2\phi=2\cos^2\phi-1, \sin2\phi=2\cos\phi\sin\phi$ this can be equivalently be rewritten as $F(\phi)=a'+b'\cos\phi+c'\sin\phi+d'\cos^2\phi+e'\sin^2\phi+f'\cos\phi\sin\phi$ with some coefficients $a',\dots, f'$. 

The average of $F(\phi)$ over $\phi\in [0,2\pi]$ is simply $\ex[F]=a$. In the single-variable case, Lemma~\ref{lemma avg} is then equivalent to the statement that $4a=F(0)+F(\pi/2)+F(\pi)+F(3\pi/2)$ and can be readily verified.

The multi-variable case is essentially the same. We call $F(\pmb\phi)$ a trigonometric polynomial of degree $2$ with respect to each variable, if Eq.~\eqref{def deg2} holds for every component of the parameter vector $\phi_k\in\pmb\phi$ (the coefficients of the expansion can depend on the rest of components in $\pmb\phi$, e.g. $F=\cos 2\phi_1 \cos 2\phi_2$ is still of degree 2). The averaging over $\pmb\phi$ is equivalent to sequential averaging over its components, and Lemma~\ref{lemma avg} follows.

\subsection{Loss function} \label{app trig}

When a generator of the parameterized gate $e^{-iG\phi/2}$ squares to identity $G^2=\mathbb{I}$, the unitary of the quantum circuit satisfies $U(\phi)=\cos{\frac{\phi}{2}}U_0+\sin\frac{\phi}{2}U_1$ with $U_0=U(\phi=0), U_1=U(\phi=\pi)$. The loss function \eqref{loss def} then has the form 
\begin{multline}
 L(\phi)=\cos^2\frac{\phi}{2} \operatorname{Tr}[\rho\,\, U_0 O U_0^\dagger]+\sin^2\frac{\phi}{2} \operatorname{Tr}[\rho\,\, U_1 O U_1^\dagger]+\\\cos\frac{\phi}{2}\sin\frac{\phi}{2} \operatorname{Tr}[\rho\,\, (U_0 O U_1^\dagger+U_1 O U_0^\dagger)]   \ .
\end{multline}
Equivalently, it can be written as $L(\phi)=a+b\cos\phi+c\sin\phi$. Hence, $L(\phi)$ is a trigonometric polynomial of degree one, and $L^2(\phi)$ of degree two, and both satisfy the assumptions of Lemma~\ref{lemma avg}.

\subsection{Loss function average in \CP{} VQA} \label{app avg}
Here we argue, that assuming $\ex[L]=0$ in a \CP{} VQA causes no loss of generality. While not necessary for our main argument, this fact simplifies the exposition and further clarifies the structure of loss function in \CP{} VQA. 

We will show, that loss function $L_i$ for any individual Pauli observable $P_i$ either (i) is constant $L_i(\pmb\phi)=\text{const}$ or (ii) has zero average $\ex[L_i]=0$. The Pauli observables with constant loss functions can be simply ignored, while the rest have zero expectation value.

To show that $L_i$ is either constant or have zero average, it is convenient to transform the original circuit. All the constant Clifford gates can be commuted through the Pauli rotations to the end of the circuit, and then absorbed into a redefined observable $O'$. Hence, the circuit can be written as $U'(\pmb\phi)=\prod_k e^{-iP'_k\phi_k/2}$ (generators $P'_k$ are in general different from the generators of the original circuit $P_k$, because of the latter are transformed during the movement of the Clifford gates). 

Now, there are two possibilities. The first is that all generators $P'_k$ commute with the observable $O'$, and therefore the corresponding loss function is trivially constant. The second case is when there exists some $P'_{k}$ anti-commuting with $O'$. Let $P'_{r}$ be the first such gate in the Heisenberg picture, i.e.
\begin{align}
    L(\pmb\phi)=\operatorname{Tr}\left[\rho\,\, e^{iP'_1\phi_1/2}\cdots e^{iP'_r\phi_r/2} O' e^{-iP'_r\phi_r/2}\cdots e^{iP'_1\phi_1/2}\right] \ .
\end{align}
Then, 
\begin{multline}
e^{iP'_r\phi_r/2} O' e^{-iP'_r\phi_r/2} = e^{iP'_r\phi_r} O'=\\\cos \phi_r O'+i P'_rO'\sin\phi_r   
\end{multline}
Therefore, the average of $L(\pmb\phi)$ with respect to $\phi_r$ vanishes, and so does the joint average with respect to all angles $\pmb\phi$.

\subsection{Fourier expansion} \label{app fourier}
The full trigonometric polynomial of the loss function is in fact the same object as its Fourier expansion~\cite{Schuld2020}, that recently has been studied from a number of angles~\cite{GilVidal2020, Atchade-Adelomou2023, Casas2023, Schreiber2022, Landman2023, Fontana2022, Fontana2022a, Nemkov2023a, Fontana2023}. 

The full Fourier expansion can be organized in levels, where level $l$ only features terms that contain $l$ angles, e.g. $\cos\phi_1\cdots\cos\phi_l$. A general Fourier series for a VQA that has $m$ parameters can contain up to $3^m$ terms, for \CP{} VQA with a single Pauli observable the upper bound is $2^m$ terms~\cite{Nemkov2023a}. A typical VQA contains exponentially many terms in its Fourier expansion, but with the exponent less than two. For instance, for circuits with random non-local observables, there are $2^{-n}\left(\frac32\right)^m$ terms on average.
\section{Exceptions} 
\subsection{No null directions} \label{app null}
\begin{figure}
\setkeys{Gin}{width=\linewidth}
        \begin{subfigure}[t]{0.20\columnwidth}
        \includegraphics{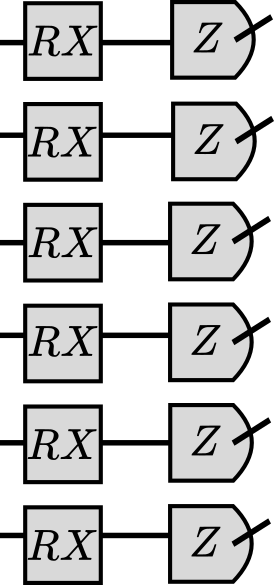}
        \caption{}
        \label{fig allx}
        \end{subfigure}\hfill%
        \begin{subfigure}{0.7\columnwidth}
        \centering
        \includegraphics{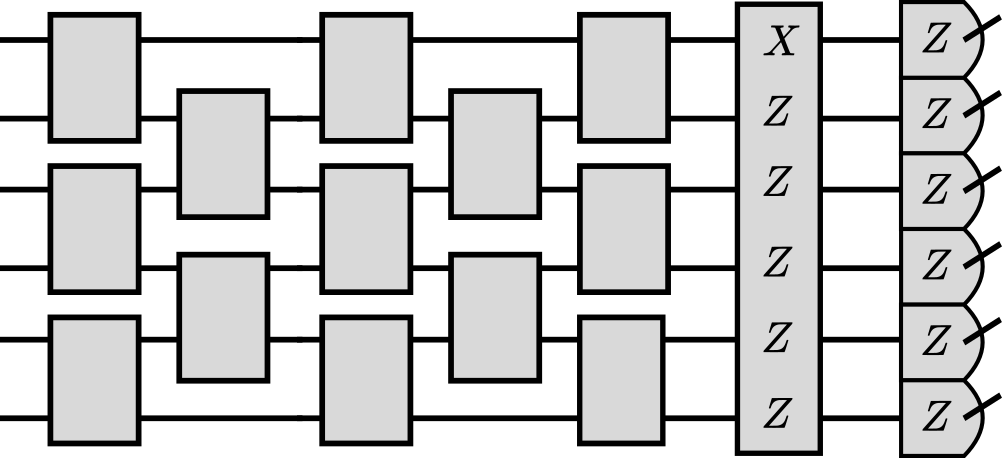}
        \caption{}
        \label{fig global}
        \end{subfigure}
    \caption{(a) A VQA with no null directions at critical points. (b) A VQA where the presence of BP depends on the state of a single gate.}
    \label{fig examples}
\end{figure}

A simple example of a VQA that does not admit null directions at the Clifford points is depicted in Fig.~\ref{fig allx}. It contains a single-qubit Pauli rotation on each qubit $U(\pmb\phi)=R_X(\phi_1)\otimes\cdots \otimes R_X(\phi_n)$, and the observable is $O=Z_1\cdots Z_n$. The loss function is simply $L=\prod_{k=1}^n \cos\phi_k$, and all local extrema correspond to $\phi_k=0, \pi$, where all $|\cos\phi_k|=1$ and hence $L=\pm1$. Perturbing any of the angles shifts away from a critical point.

This example, however, appears to be rather contrived. For one, it features a specially chosen global observable. More importantly, the Fourier expansion of this loss function looks very atypical -- it contains a single term at the largest level. Each term in the Fourier expansion corresponds to a Clifford point, where the value of the loss function equals $\pm1$. The number of null directions at a Clifford point, corresponding to a term at level $l$, is simply $m-l$. It was demonstrated in Ref.~\cite{Nemkov2023a} that the distribution of terms by level in a Fourier expansion typically shows a bell-type curve, centered around $m_0=m/2$ for non-local loss, and at $m_0\le m/2$ for local loss functions.  As argued in Sec.~\ref{sec disc} we generically expect at least $m/2$ null directions, the same estimate that comes from the distribution of terms in the Fourier expansion. While we do not rule out the existence of interesting examples of VQA having most of the Fourier terms clustered at high levels, they are certainly not generic, and do not include e.g. QAOA or hardware-efficient circuits~\cite{Nemkov2023a}.

\subsection{Barren plateau from a single gate} \label{app bp der}
A VQA in Fig.~\ref{fig global} gives a simple yet contrived example, where switching a single gate can eliminate the BP. The first part of the circuit is assumed to be local and shallow, while the last gate is a global Pauli rotation with generator $P=X_1Z_2\cdots Z_n$. The observable $O=Z_1\dots Z_n$ is also non-local. When the angle of the global Pauli gate is zero, the VQA has a BP due to non-locality of the observable. However, choosing the angle $\phi=\frac{\pi}{2}$, so that the Pauli rotation gate is $e^{-iP \frac{\pi}{4}}$, effectively removes the non-locality of the observable since $e^{iP \frac{\pi}{4}}Oe^{-iP \frac{\pi}{4}}=e^{iP \frac{\pi}{2}}O=iPO=Y_1$. Hence, the resulting VQA is not expected to have a BP.

\subsection{Variance of a loss function with a random observable} \label{app random}
Here we prove relation \eqref{expvar}. First, note that inequality $\var[L_P]=\ex[L_P^2]-\ex[L_P]^2 \le \ex[L_P^2]$ holds for any random variable $L_P$. Next, according to \eqref{var as sum}, the expectation value $\ex[L_P^2]$ can be represented as a sum over Clifford points of the form 
\begin{align}
    L_P^2(\pmb\phi_c)=\langle 0| U(\pmb\phi_c)^\dagger P U(\pmb\phi_c) |0\rangle^2 \ .
\end{align}
The average value over $P$ for any choice of $\pmb\phi_c$ is simply equal to
\begin{multline}
    \ex_P[L_P^2(\pmb\phi_c)]=\ex_{P}\left[\langle 0| U(\pmb\phi_c)^\dagger P U(\pmb\phi_c) |0\rangle^2\right]=\\\ex_{P}\left[\langle 0|P|0\rangle^2\right]= 2^{-n}\ ,
\end{multline}
which is just a probability that a random Pauli operator has a non-vanishing expectation value in the all-zero state $\rho=(|0\rangle\langle 0|)^{\otimes n}$. Relation \eqref{expvar} thus follows.

\section{Details of numerical experiments} \label{app numerics}
Both experiments reported in Fig.~\ref{fig numerics} use circuit layouts specified in Fig.~\ref{fig circuit} with number of qubits varying from $n=2$ to $n=10$ and number of layers between $10$ and $50$. All circuits simulated in Fig.~\ref{fig exact} have 50 layers.

To gather data for Fig.~\ref{fig bp}, for each pair (number of qubits, number of layers) and for each observable from $\mathcal{P}_2^{NN}$ we sample 50 values of $\pmb\phi$ from a uniform distribution, and 50 Clifford values. Note that the for 10-qubit circuits at 30 and 50 layers the conditioned Clifford marker $\triangle$ is absent, because no Clifford points with more than a single non-zero Pauli appeared during sampling.

The data for Fig.~\ref{fig exact} was obtained along the lines described in the main section Sec.~\ref{sec numerics}. We are not aware of any efficient procedure that, for a given circuit and Pauli observable, finds a Clifford point where this observable has non-zero expectation. Thus, we resort to extensive sampling over Clifford points and trying many observables in parallel. For this reason, in this experiment, we choose a larger set of non-nearest neighbor Pauli strings $\mathcal{P}_2$.

For each number of qubits we perform 10 trials, each producing a potential exact LM, and compute the statistics over remaining loss function terms at these points. Each individual trial is depicted by $\bigcirc$ in Fig.~\ref{fig exact}. The average value over all trials for a given number of qubits is depicted by $\square$.

While looking for the next Pauli operator to add, we compute values of all independent Pauli operators from $\mathcal{P}_2$ at
\begin{align}
    \left\lfloor 30\frac{2^n}{|\mathcal{P}_2|}\right\rfloor \approx 3 \frac{2^n}{n^2} \label{num samples}
\end{align}
random Clifford points. We scale the number of sampled points exponentially, to keep the probability of finding non-zero expectation values high enough. If no Pauli operators with non-zero expectation values were found among the sampled points, we stop the search for additional Pauli operators. In two cases for $n=10$ qubit circuit this procedure did not produce even a single Pauli operator, and the number of individual samples depicted at Fig.~\ref{fig exact} for $n=10$ is correspondingly reduced.

This procedure ends by proposing an exact LM, optimizing several Pauli terms $P_I$ and induces a split of the parameter space $(\pmb\theta_I, \pmb\varphi)$. We then test if the loss functions, corresponding to Pauli terms, remaining in $\mathcal{P}_2$ and independent of $P_I$, are identically zero as functions of $\pmb\varphi$. We also test that their gradients with respect to $\pmb\theta_I$ are identically vanishing as well. We chose not to study Hessian entries, but expect largely similar results in this case.

As discussed in Sec.~\ref{sec numerics}, the number of gradient components for deep enough circuits is substantial. Hence, we limit the number of gradient components (chosen at random for each trial) evaluated for each observable from $\mathcal{P}_2$ by the same value featuring in Eq.~\ref{num samples}. We consider each loss function or gradient component corresponding to an observable from $\mathcal{P}_2$ to be exactly vanishing, if the variance estimated over 10 uniformly sampled points is zero within the machine precision. 

The code producing the numerical results reported here is mostly written using PennyLane library \cite{Bergholm2018}, and is available at \cite{barren_traps}.

\bibliographystyle{quantum}
\bibliography{library.bib, additional_refs.bib}

\end{document}